\documentclass[apj,square,numbers,iop]{myemulateapj}
\usepackage{graphicx}
\usepackage{epstopdf}
\usepackage{xcolor}
\usepackage{amsmath,amssymb,latexsym,bm}
\usepackage{hyperref}
\usepackage{ulem}
\usepackage{times}
\newcounter{RomanNumber}

\def\be{\begin{equation}}
\def\ee{\end{equation}}
\def\ba{\begin{eqnarray}}
\def\ea{\end{eqnarray}}
\begin{document}
\title{Revisiting the Instability Problem of Interacting Dark Energy Model in the Parametrized Post-Friedmann Framework}
\author{Ji-Ping Dai$^{1}$, Jun-Qing Xia$^{1}$}

\affil{$^{1}$Department of Astronomy, Beijing Normal University, Beijing 100875, China; xiajq@bnu.edu.cn}

\begin{abstract}
  Dark energy might interact with dark matter in a direct, non-gravitational way, which can help remedy several theoretical defects. In order to find out the properties of interacting dark energy models, it is necessary to investigate the cosmological perturbations in detail. However, due to the improper use of pressure perturbation of dark energy, a large-scale instability at the early stage occurs occasionally. In recent years, parametrized post-Friedmann approach has been used to calculate the perturbation equations in the interacting dark energy scenario. Under this framework, the dark energy pressure perturbation was replaced by the relationship between the momentum density of dark energy and other components on a large scale. However, this paper shows that if the interaction terms are related to the velocity perturbation of dark energy, the density perturbation of dark matter and the matter power spectrum will diverge when the equation of state parameter of dark energy $w$ is close to $-1$. A simple parameterization steering clear of this problem is proposed in this paper which is a more general form and can be applied to explore the interaction between dark matter and dark energy by using various cosmological data.
\end{abstract}

\maketitle

\section{Introduction}

Dark energy and dark matter are two major scientific issues in fundamental physics in the 21st century. As two dominant sources, they are indirectly detected via their gravitational effects nowadays. Except for gravitational interaction, the direct non-gravitational interaction which does not violate current observational constraints should also be taken into consideration. Such a dark sector interaction can help overcome several theoretical defects about dark energy, such as the cosmic coincidence problem \citep{Amendola2000, ZhangX2005, CaiRG2005}. A detailed discussion about this interaction can promote the understanding of dark matter and dark energy.

Human beings know very little about the nature of dark matter and dark energy, so it is hard to work out a reasonable energy transfer rate $Q$ from first principles. It is only accessible to construct some interaction models phenomenologically, such as $Q= 3\beta H \rho_c$ and some more complicated models \citep{SkordisC2015}. In order to find out a better model that can reflect the real process of interaction, it is necessary to constrain the extra parameters resulting from interacting dark energy model by using the latest cosmological data.

In order to testify the assumption of interaction between dark matter and dark energy and  constrain the intensity of interaction, Einstein and fluid equations deserve serious reconsideration. In an interacting dark energy (IDE) scenario, the energy conservation equations of dark energy and cold dark matter can be written as,
\begin{subequations}
\label{rhodec}
\begin{align}
&\rho_{de}' = -3\mathcal{H}(1+w)\rho_{de}+aQ_{de} \ ,\\
&\rho_c' = -3\mathcal{H}\rho_c+aQ_c \ ,
\end{align}
\end{subequations}
where $Q_{de} = -Q_c = Q$ refers to the energy transfer rate, the prime represents derivative with respect to the conformal time $\eta$, $\rho_{de}$ and $\rho_c$ are energy densities of dark energy and cold dark matter, respectively. $\mathcal{H} = a'/a$ is the conformal Hubble expansion rate, $a$ is the scale factor of the universe, and $w$ is the equation of state parameter of dark energy. There are several works that have been done to constrain $Q$ by using cosmic microwave background (CMB), the baryon acoustic oscillation (BAO) and the type Ia supernovae (SNIa) \citep{GuoZK2007, BohmerCG2008, KoyamaK2009, HeJH20092, HeJH20093, XiaJQ2009, MartinelliM2010, WeiH2011, LiYH2014, YangWQ2014}.

However, the above results cannot be used to explore the full parameter space due to the well-known large-scale instability existing in the IDE scenario, which behaves as the blow up of the curvature perturbation on a large scale for some specific values of the dark energy equation of state $w$ and the coupling constant $\beta$ \citep{MajerottoE2008}. So in order to avoid the instability, it should be assumed that $w > -1$ and $\beta>0$ for $Q\propto \rho_{de}$ models \citep{HeJH2009, ClemsonT2012} or $w<-1$ for $Q\propto \rho_{c}$ models \citep{MajerottoE2008}.

To use fewer parameters, people often consider dark energy as non-adiabatic fluid, which can cause the divergence problem when $w$ crosses the phantom divide $w = -1$ \citep{AlexanderV2005, CaldwellRR2005, HuW2005, ZhaoGB2005}. Wayne Hu came up with an effective framework: parametrized post-Friedmann (PPF) approach to solve the divergence problem \citep{HuW2008, FangW2008}. In this framework, perturbation equations do not contain the perturbed variables of dark energy which will be calculated in other equations. In recent years, PPF framework has been used in the IDE scenario \citep{LiYH20141,LiYH20142}. It can constrain the interacting dark energy models without assuming any specific priors on $w$ and $\beta$. And the fit results show that the PPF framework is effective in exploring the full parameter space of interacting models.

However, the previous works only constrain the interacting models like $Q_{de}^{\mu} \propto u_c^{\mu}$, where $u_c^{\mu}$ is four velocity of dark matter. In this paper, the PPF framework for the IDE scenario will be studied intensively. The result shows that when $w$ is close to $-1$, the velocity perturbations of dark energy in synchronous gauge, $v_{de}$ will diverge. So if the perturbation of interaction is proportional to $u_{de}^{\mu}$, the perturbation system will  diverge, including matter power spectrum. Therefore, another interacting dark energy model is needed to avoid this problem.

This paper is organized as follows. In Sec.\ref{sec2}, the general perturbation equations in terms of four perturbation variables and conservation equations in the IDE scenario will be calculated, and then the large-scale instability in Newtonian gauges will be reviewed. In Sec.\ref{sec3}, the PPF approach in comoving gauge established in the previous paper \citep{LiYH20141} will be introduced and then another divergence when $w$ close to $-1$ in PPF framework will be discussed in Sec.\ref{sec4}. In the meantime, a new interacting dark energy model to steer clear of the divergence will be put forward. The conclusion is in Sec.\ref{sec5}.

\section{Perturbation equations and large-scale instability of interacting dark energy model}
\label{sec2}
\subsection{Perturbation equations in the IDE scenario}
Scalar perturbation plays the most important role in the interaction between dark matter and dark energy. Under a FRW universe, the scalar metric perturbations can be expressed in general with four functions, $A, B, H_L$ and $H_T$ \citep{KodamaH1984, Robert2003}.
\begin{subequations}
\label{4raodong}
\begin{align}
\delta g_{00} &= -2a^2A \ , \\
\delta g_{0i} &= -a^2B_{,i} \ , \\
\delta g_{ij} &= a^2(2H_L\delta _{ij}+2D_{ij}H_{T}) \ ,
\end{align}
\end{subequations}
where $B_{,i} = \partial B /  \partial$$x^i$, $D_{ij}=(\partial_i\partial_j-\frac{1}{3}\delta_{ij}\nabla^2)$. Similarly, the perturbation of the stress-energy tensor can be expressed by another four functions.
\begin{subequations}
\label{ttttt}
\begin{align}
\delta T^0_{~0} &= -\delta \rho \ , \\
\delta T^i_{~0} &= -(\bar \rho+\bar p) \partial^i v \ ,\\
\delta T^i_{~j} &= \delta p\delta^i_j+\Pi^i_j \ .
\end{align}
\end{subequations}
$\delta \rho$ is the energy density perturbation, $\partial^i v$ refers to velocity perturbation, and $\delta p$, $\Pi^i_j$ are pressure perturbation and anisotropic stress perturbation, respectively. Therefore, the general four perturbed Einstein field equations can be derived by using perturbed scalar metric and stress-energy tensor.
\begin{subequations}
\label{kst}
\begin{align}
\begin{split}
&-k^2 H_L + 3 \mathcal{H} (\mathcal{H} A- H_L')- \mathcal{H} k B - \frac{1}{3} k^2 H_T  \\ \label{firEin}
& = -4\pi G a^2 \delta \rho \ ,\\
\end{split}
\\
\begin{split}
-k \mathcal{H} A + k H_L'+ \frac{1}{3}k H_T'= -4 \pi G a^2(p+\rho)(v-B),\\ \label{secEin}
\end{split}
\\
\begin{split}
&-k^2(H_L+A) + H_T''+ 2\mathcal{H}H_T'-kB'-2 \mathcal{H} k B -\frac{1}{3}k^2H_T\\
& = 8\pi G a^2 \Pi \ ,\\
\end{split}
\\
\begin{split}
&\mathcal{H} A' +(2 \mathcal{H}'+\mathcal{H}^2) A-\frac{1}{3}k ^2(A+H_L)-2\mathcal{H}H_L'\\
&-H_L''+\frac{2}{3}\mathcal{H}kB +\frac{1}{3}k B'-\frac{1}{9}k^2H_L =  4 \pi G a^2 \delta p \ .
\end{split}
\end{align}
\end{subequations}
All the equations have been converted into momentum space with $k$ the wave number.

The above equations are calculated without considering the interaction between dark matter and dark energy. In the IDE scenario, the results are the same because the stress-energy tensor contains all the components of the universe. As for conservation equations, in order to calculate the fluid equations for each part of the universe, the results have to be modified. In the IDE scenario, conservation law is $\nabla _{\nu}T^{\mu \nu}_{I}=Q^{\mu}_I$, $\sum_I Q^{\mu}_I=0$, where $Q^{\mu}_I$ is the energy-momentum transfer vector of $I$ fluid. It can be written in a general form \citet{KodamaH1984,MajerottoE2008},
\begin{equation}
\label{bijiao}
Q^I_{\mu}=a(-Q_I(1+A)-\delta Q_I, [f_I + Q_I(v - B)]_{,i}) \ ,
\end{equation}
where $\mu=0$ and $\mu=i$ refer to the energy transfer and the momentum transfer of $I$ fluid. For $\nu = 0$, the conservation laws become continuity equations; and for $\nu = i$, the conservation laws become Euler equations \citep{KodamaH1984},
\begin{subequations}
\label{CNS}
\begin{align}
\begin{split}
&\delta \rho_I' + 3\mathcal{H}(\delta \rho_I+\delta p_I)+(\rho_I+p_I)(3H_L'+kv_I)\\
&=a(\delta Q_I + AQ_I)\label{con} \ ,\\
\end{split}
\\
\begin{split}
&[(\rho_I+P_I)(v_I-B)]'+4\mathcal{H}(\rho_I+P_I)(v_I-B)+\frac{2}{3}kc_K p_I\Pi\\
&-k [\delta P_I +(\rho_I + P_I)A] =a( f_I + Q_I(v - B))\label{eulur} \ ,
\end{split}
\end{align}
\end{subequations}
where $c_K=1-3K/k^2$ with $K$ the spatial curvature. In this paper a flat universe is assumed, so $K=0$ and $c_K=1$.

The covariant Einstein and conservation equations can be applied to any choice of gauge. Under gauge transformation, only two functions are needed to specify a gauge. For more details about the gauge transformation, please refer to \citet{Robert2003}.

Three useful gauges are used in this paper.
\begin{itemize}
  \item Newtonian gauges, which is specified by the conditions \citep{MaCP1995},
  \begin{equation}
  B = 0 ,H_T =0 ,\Psi \equiv A ,\Phi \equiv -H_L \ .
  \end{equation}
  \item Comoving gauge, which is useful when constructing the PPF description \citep{HuW2008},
  \begin{equation}
  B = v ,H_T = 0 ,\xi \equiv A ,\zeta \equiv H_L \ .
  \end{equation}
  \item Synchronous gauge, which is used in numerical calculation \citep{MaCP1995},
  \begin{equation}
  A = 0 ,B = 0 ,\eta_T \equiv -\frac{1}{3} H_T-H_L ,h_L \equiv 6 H_L \ .
  \end{equation}
\end{itemize}

It can be found that there are six variables and only four equations are independent. The relationships between $\delta p, \Pi$ and the density fluctuations need to be built to complete the equations. Generally, people construct the relationship between pressure and density fluctuations by defining the sound speed $c_s$ and adiabatic sound speed $c_a$.
\begin{equation}
\label{yaqiang}
\delta p = c_a^2\delta \rho+(c_s^2-c_a^2)(\delta \rho - \bar \rho'\frac{v-B}{k}) \ .
\end{equation}
After putting this relationship into Eq.\ref{CNS}, the equation will be,
\begin{subequations}
\label{DV}
\begin{align}
&\delta_I'+3 \mathcal{H} (c^2_{s,I}-w_I)\delta_I  +9\mathcal{H}^2(c^2_{s,I}-c^2_{a,I})(1+w_I)\frac{v_I-B}{k}\nonumber\\
&+3(1+w_I)H_L'+ (1+w_I)kv_I\nonumber\\
&=\frac{aQ_I}{\rho_I}\left[A-\delta_I+3\mathcal{H}(c^2_{s,I}-c^2_{a,I})\frac{v_I-B}{k}\right]+\frac{a}{\rho_I}\delta Q_I \ ,\\
&(v_I-B)' + \mathcal{H}(1-3c^2_{s,I})(v_I-B))-\frac{c^2_{s,I}}{1+w_I}k\delta_I-kA\nonumber\\
&=\frac{aQ_I}{(1+w_I)\rho_I}\left[{v-B}-(1+c^2_{s,I})(v_I-B)\right]+\frac{af_I}{(1+w_I)\rho_I} \ ,
\end{align}
\end{subequations}
where $\delta_I = \delta \rho_I/\rho_I$.

\subsection{Large-scale instability of interacting dark energy model in Newtonian gauges}
The rest part of this section mainly talks about the large-scale instability of interacting dark energy model. For the model described by the covariant energy-momentum transfer four-vector $Q_c^{\nu} = -Q_{de}^{\nu}=-3\beta H\rho_cu^{\nu}_c$, the transfer terms can be derived as,
\begin{subequations}
\label{previous}
\begin{align}
&Q_{de} = -Q_c = 3\beta H\rho_c\ ,\\
&\delta Q_{de} = -\delta Q_c = 3\beta H\rho_c\delta_c \ ,\\
&f_{de}= -f_c=3\beta H\rho_c(v_c-v) \ .
\end{align}
\end{subequations}
In Newtonian gauges, the density and velocity perturbation equations of dark energy and dark matter Eq.\ref{DV} can be written as,
\begin{subequations}
\begin{align}
&\delta'_{de}  + (1+w)(kv_{de}-3\Phi') + 9\mathcal H^2(1-w^2)\frac{v_{de}}{k}\nonumber\\
&+ 3\mathcal H(1-w)\delta_{de}= 3a\beta H \frac{\rho_c}{\rho_{de}}\left[\delta_c -\delta_{de} + 3\mathcal H (1-w)\frac{v_{de}}{k}+\Phi\right] \ ,\\
&v_{de}' -2 \mathcal H v_{de} -\frac{k}{(1+w)}\delta_{de} -k\Phi =\frac{3a\beta H}{(1+w)}\frac{\rho_c}{\rho_{de}}\left(v_c-2v_{de} \right),\\
&\delta'_c +kv_c -3\Psi' =- 3a\beta H\Phi \ ,\\
&v_c' + \mathcal H v_c -k\phi = 0 \ .
\end{align}
\end{subequations}
Here, set $c^2_{s,de} = 1$.

Now consider the early radiation era. In this period, $\mathcal H = \eta^{-1}$. So the four perturbed Einstein field equations in Newtonian gauges become,
\begin{subequations}
\begin{align}
&k^2 \Phi + 3 \eta^{-1}\Phi' +3\eta^{-2}\Psi' = -4\pi G a^2 \delta \rho \ ,\\
&k \eta^{-1} \Psi + k \Phi'   = 4 \pi G a^2(p+\rho)v \ ,\\
&k^2(\Phi-\Psi)= 8\pi G a^2 \Pi_{\nu} \ ,\\
&\Phi'' + \eta^{-1} \Psi' + 2 \eta^{-1} \Phi' - \eta^{-2} \Psi' + \frac{1}{3}k ^2(\Phi-\Psi)=  4 \pi G a^2 \delta p \ .
\end{align}
\end{subequations}
In order to work out a solution in the super-Hubble scale limit, $k\eta \ll 1$, a leading-order power-law form for the perturbations is assumed,
\begin{equation}
\begin{split}
&\Phi  = A_{\Phi} (k\eta)^{n_{\Phi}} ,\Psi  = A_{\Psi} (k\eta)^{n_{\Psi}} ,\\
&\delta_I  = B_{I} (k\eta)^{n_{I}} ,kv_I  = C_{I} (k\eta)^{s_{I}} \ .
\end{split}
\end{equation}
The two first order differential equations of dark energy can be converted to one second order differential equation, and then get the solution of $n_\Phi$ \citep{MajerottoE2008},
\begin{equation}
n_{\Phi} = \frac{-(1+2w)\pm \sqrt{3w^2-2}}{1+w} \ .
\end{equation}
It can be figured out that when $-1 < w < -\sqrt{\frac{2}{3}}$, $n_{\Phi}>0$, the curvature perturbation will continue to grow, which results in large-scale instability.

The large-scale instability seriously hinders the studies of IDE. Until this problem is solved, people have to investigate IDE models in part of their parameter space. In the next section, an attempt to avoid this instability will be introduced to solve the divergence problem when $w$ cross $w=-1$.

\section{PPF framework for the IDE scenario}
\label{sec3}
Obviously, the large-scale instability is caused by improper use of calculation of $\delta p_{de}$. In the normal case, dark energy is considered as non-adiabatic fluid. The interaction between dark matter and dark energy will lead to the rapid growth of non-adiabatic mode. It is necessary to find a more effective framework to calculate the cosmological perturbations of dark energy. In the previous work \citep{LiYH20141}, they established a PPF framework for the IDE scenario, which can successfully solve the problem of instability in the IDE models.

In this paper, the PPF framework in comoving gauge is applied with the symbols of \citet{LiYH20141}, $\zeta \equiv H_L$, $\xi \equiv A$,$\rho \Delta \equiv \delta \rho$, $\Delta p \equiv \delta p$, $V \equiv v$, $\Delta Q\equiv \delta Q$. $B=V_T$ where $V_T$ represents the velocity perturbation of total matters except dark energy. In the following part of this paper, the subscript $T$ is used to refer to all the components without dark energy.  $\Pi$ and $f_I$ are gauge invariant, so the expressions remain unchanged. For dark matter,  $\Delta p_c = \Pi_c = 0$, the fluid equations can be calculated by Eq.\ref{DV}. As for dark energy, $\Pi_{de}$can be set as 0 and another equation is needed to complete the dark energy perturbation system.

Under PPF framework, the additional equation should satisfy two requirements. The first one is on superhorizonal scales. When $k_H = k/\mathcal{H} \ll 1$, the relationship between $V_{de}-V_T$ is parameterized by a function $f_\zeta(a)$ \citep{HuW2008, FangW2008}.
\begin{equation}
\lim_{k_H \ll 1}{4\pi G a^2 \over \mathcal{H}^2} (\rho_{de} + p_{de}) {V_{de} - V_T \over k_H}
= - {1 \over 3}   f_\zeta(a) k_H V_T \ .
\end{equation}
Putting this relationship into Eq.\ref{secEin} in a flat universe, and the equation of motion for the curvature perturbation $\zeta$ on a large scale can be obtained,
\begin{equation}
\label{need2}
\lim_{k_H \ll 1}{\zeta'} = \mathcal{H}\xi + {1 \over 3} f_\zeta(a) k V_T \ ,
\end{equation}
where $\xi$ in comoving gauge can be obtained from Eq.\ref{CNS},
\begin{equation}
\label{need3}
\xi =  -{\Delta p_T - {2\over 3}p_T\Pi_T + {a \over k} [Q_c(V-V_T)+f_c] \over \rho_T + p_T} \ .
\end{equation}

The second condition is on the small scales $k_H \gg 1$. The evolution of the curvature perturbation is described by the Poisson equation,$\Phi = 4\pi G a^2 \rho_T\Delta_T / k^2$, where $\Phi$ is the perturbation of matric in Newtonian gauges. Then a dynamical function $\Gamma$ will be introduced to meet the two conditions,
\begin{equation}
\label{need1}
\Phi + \Gamma = \frac{4\pi Ga^2}{k^2}\rho_T\Delta_T \ .
\end{equation}

In order to complete the equations, it is necessary to figure out the differential equation of $\Gamma$. Take the derivative of both sides of Eq.\ref{need1} with respect to the conformal time, together with Eq.\ref{firEin} and the process can be written as follows,
\begin{equation}
\begin{split}
 &\Gamma ' = -\Phi' + 2 \mathcal{H} \frac{4\pi Ga^2}{k^2}(\rho_T\Delta_T)-a(Q_c\xi+\Delta Q_c)] \\
 &- \frac{4\pi Ga^2}{k^2}[3\mathcal{H}( \rho_T \Delta+\Delta p_T)+(\rho_T+P_T)(3\zeta'+kV_T)\\
 &=-\Phi'-\frac{4\pi Ga^2}{k^2}3\mathcal{H}\Delta p_T-\frac{4\pi Ga^2}{k^2}(\rho_T+p_T)kV_T\\
 &+\frac{4\pi Ga^2}{k^2}[-3\zeta'(\rho_T+p_T)+a(\Delta Q_c+\xi Q_c)]\\
 &-\mathcal{H} \frac{4\pi Ga^2}{k^2}(\rho_T\Delta_T) \ .
\end{split}
\end{equation}
Then with the help of Eq.\ref{need2} and Eq.\ref{need3}, $\Gamma'$ at $k_H \ll 1$ can be written as,
\begin{equation}
\begin{split}
 \lim_{k_H \ll 1}&{\Gamma '} =-\Phi'- \mathcal{H} \frac{4\pi Ga^2}{k^2}(\rho_T\Delta_T)-\frac{4\pi Ga^2}{k^2}(\rho_T+p_T)kV_T\\
 &-\frac{8\pi Ga^2}{k^2}\Pi_T+\frac{4\pi Ga^2}{k^2}\left\{-f_{\zeta}(\rho_T+p_T)kV_T \right.\\
 &\left.+{3a\over k_H}[Q_c(V-V_T)+f_c]+a(\Delta Q_c+\xi Q_c)\right\} \ .
\end{split}
\end{equation}
Here Einstein equations under Newtonian gauge and the gauge transformation, $\Psi = \xi +V_T'/k + V_T/k_H$, $\Phi = \zeta + V_T/k_H$ can be used to get the following equation,
\begin{equation}
\label{gamma}
 \lim_{k_H \ll 1}{\Gamma '}=S - \mathcal{H}\Gamma \ ,
\end{equation}
  which is the same as the case when there is no interaction between dark matter and dark energy \citep{HuW2008}. However, if there is no interaction,
\begin{equation}
 S = \frac{4\pi Ga^2}{k^2}\left\{[(\rho_{de}+p_{de})-f_{\zeta}(\rho_T+p_T)]kV_T\right\} \ .
\end{equation}
In the IDE scenario, $S$ may be modified as,
\begin{equation}
\begin{split}
 &S = \frac{4\pi Ga^2}{k^2}\left\{[(\rho_{de}+p_{de})-f_{\zeta}(\rho_T+p_T)]kV_T\right.\\
 &\left.+{3a\over k_H}[Q_c(V-V_T)+f_c]+a(\Delta Q_c+\xi Q_c)\right\} \ ,
 \end{split}
\end{equation}
which may cause system divergence when $w$ is close to $-1$(see below).

From Eq.\ref{need1}, it can be found that $\Gamma\rightarrow 0$ when $k_H \gg 1$. With a transition scale parameter $c_\Gamma$, the equation of motion for $\Gamma$ on all scales can be written as,
\begin{equation}
\label{gammaequ}
 (1+c^2_{\Gamma}k^2_{H})[\Gamma ' + \mathcal {H} \Gamma + c^2_{\Gamma}k^2_{H}\mathcal{H}\Gamma]=S \ .
\end{equation}

Once the evolution of $\Gamma$ is obtained, it is accessible to get the energy density and velocity perturbations of dark energy,
\begin{equation}
\label{dev}
 \begin{split}
 &V_{de}-V_T = \frac{-k}{4\pi Ga^2(\rho_{de}+P_{de})F} \left[ S-\Gamma' - \mathcal{H}\Gamma \right.\\
 &\left.+ f_{\zeta}\frac{4\pi Ga^2(\rho_{de}+P_{de})}{k}V_T \right] \ .
\end{split}
\end{equation}
\begin{equation}
\label{derho}
 \rho_{de}\Delta_{de} = -3(\rho_{de}+P_{de})\frac{V_{de}-V_T}{k_H}-\frac{k^2\Gamma}{4\pi G a^2} \ ,
\end{equation}
with $F=1+12\pi G a^2(\rho_T+p_T)/k^2$. In the end, it is necessary to give the forms of functions of $f_{\zeta}$ and $c_{\Gamma}$. For the value of $c_\Gamma$, the perturbation evolutions of dark energy are insensitive to its value according to \citet{FangW2008}. We choose it to be 0.4 in this paper. For $f_{\zeta}$, it suffices for most purposes to simply let $f_{\zeta} = 0$ \citep{FangW2008}. To get the dark energy perturbations in the synchronous gauge, please refer to \citep{HuW2008, LiYH20142}

\section{System divergence when $w$ close to $-1$ in PPF framework}
\label{sec4}
\subsection{Divergence of  velocity perturbations of dark energy}
It seems that the problem of the large-scale instability in all the IDE models can be successfully solved within such a generalized PPF framework. In \citet{LiYH20141} they constrain the $Q_{de}^{\mu}=3\beta H\rho_c u_c^{\mu}$ model. And in \citet{LiYH20142} they constrain the $Q_{de}^{\mu}=3\beta H\rho_{de} u_c^{\mu}$ model by using the iteration method. They also testify the models of vacuum energy interacting with cold dark matter in \citet{LiYH2015}by using PPF approach. However, there is still a minor problem of this method.

\begin{figure*}[tbp]
\includegraphics[scale=0.65]{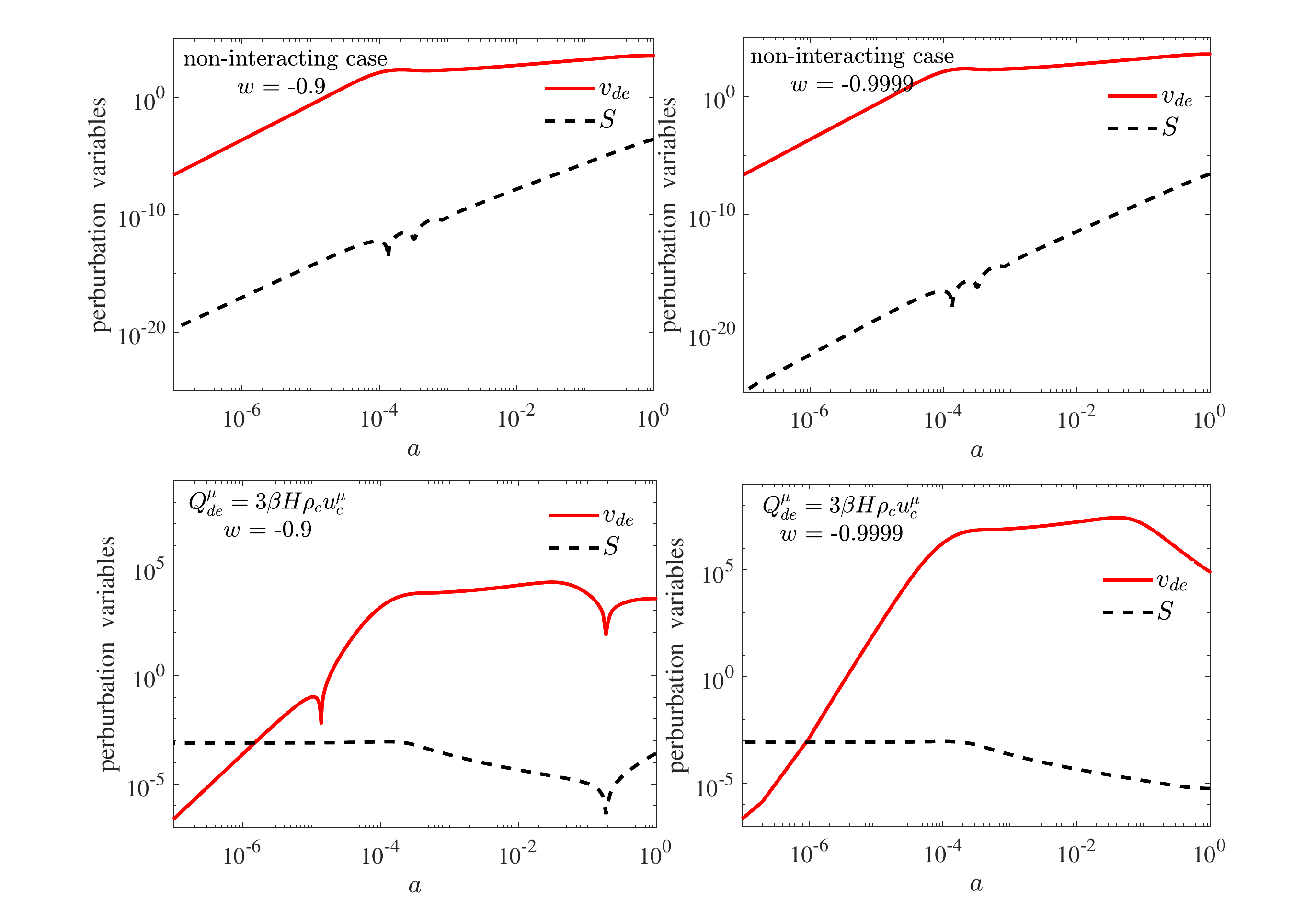}
\caption{$v_{de}$ (red solid line) and $S$ (black dashed line) under synchronous gauge. The upper panels are obtained by non-interacting case, while the lower panels use $Q_{de}^{\mu}=3\beta H\rho_c u_c^{\mu}$ model, with $w=-0.9$, $k=0.1\rm Mpc^{-1}$}
\label{fig1}
\end{figure*}

It can be found that Eq.\ref{dev} is invariant under gauge transformation and the right side of this equation is proportional to $1/(1+w)$, which should come into notice when $w$ is close to $-1$. In the case that there is no interaction between dark matter and dark energy, we have $S-\Gamma'-\mathcal{H}\Gamma = 0$ when $k_H \ll 1$ and  $S-\Gamma'-\mathcal{H}\Gamma = S$ when $k_H \gg 1$. Eq.\ref{gamma} shows that $S$ is proportional to $\rho_{de}(1+w)$ if there is no interaction. So $V_{de}-V_T$ does not diverge when $w$ is very close to $-1$. In the upper two figures of Fig.\ref{fig1}, we plot $v_{de}$ which transform form $V_{de}$ under comoving gauge into synchronous gauge and $S$ under synchronous gauge, with $w = -0.9$ and $w=-0.9999$, respectively. The wave number $k=0.1\rm Mpc^{-1}$. If $w$ is very close to $-1$, $S$ will be close to $0$, and therefore, $v_{de}$ is slightly changed by the value of $w$.

However, in the the interacting dark energy model, $S$ is modified and has an additional part ${3a\over k_H}[Q_c(V-V_T)+f_c]+a(\Delta Q_c+\xi Q_c)$ which is not proportional to $1+w$. So $v_{de}$ would diverge when $w$ is very close to $-1$. For example, when using $Q_{de}^{\mu}=3\beta H\rho_c u_c^{\mu}$ model, $Q_c = -3 \beta H \rho_c, \delta Q_{c}= -3\beta H\rho_c\delta_c, f_c = -3\beta H \rho_c(v_c-v)$. We set $k=0.1\rm Mpc^{-1}$, $\beta = -0.001$ which is favoured by \citet{LiYH20141}, $w=-0.9$ and $w=-0.9999$, respectively.  The lower figures of Fig.\ref{fig1} show the results.

The result shows a totally different situation, because $S$ in the IDE scenario has an extra part that is not proportional to $1+w$. When $w$ is close to $-1$, $v_{de}$ would diverge. What is more, $v_{de} - v_T$ is a gauge invariant variable, so it cannot be avoided by gauge transformation.

However, the system would not diverge unless the interacting dark energy model is related to $v_{de}$. For example, if the model $Q_{de}^{\mu}=3\beta H\rho_c u_c^{\mu}$ is used, the interaction only affects the evolution of dark matter and dark energy, and the perturbed continuity and Euler equations Eq.\ref{DV} of dark matter can be reduced to,
\begin{subequations}
\label{derho}
\begin{align}
&\delta_c' + kv_c + 3H_L' = -3aH\beta A, \\
&(v_c-B)'+\mathcal{H}(v_c-B) -kA = 0 .
\end{align}
\end{subequations}
It can be found that the Euler equation is the same as that in the non-interacting case, which means there is no violation of weak equivalence. Under synchronous gauge, $v_c$ can be written as $0$. The only signal of the dark sector interaction in the structure formation to linear order is via the modification of the background expansion history \citep{KazuyaK2009}. So the divergence of $v_{de}$ does not impact the evolution equations.

\begin{figure*}[tbp]
\includegraphics[scale=0.67]{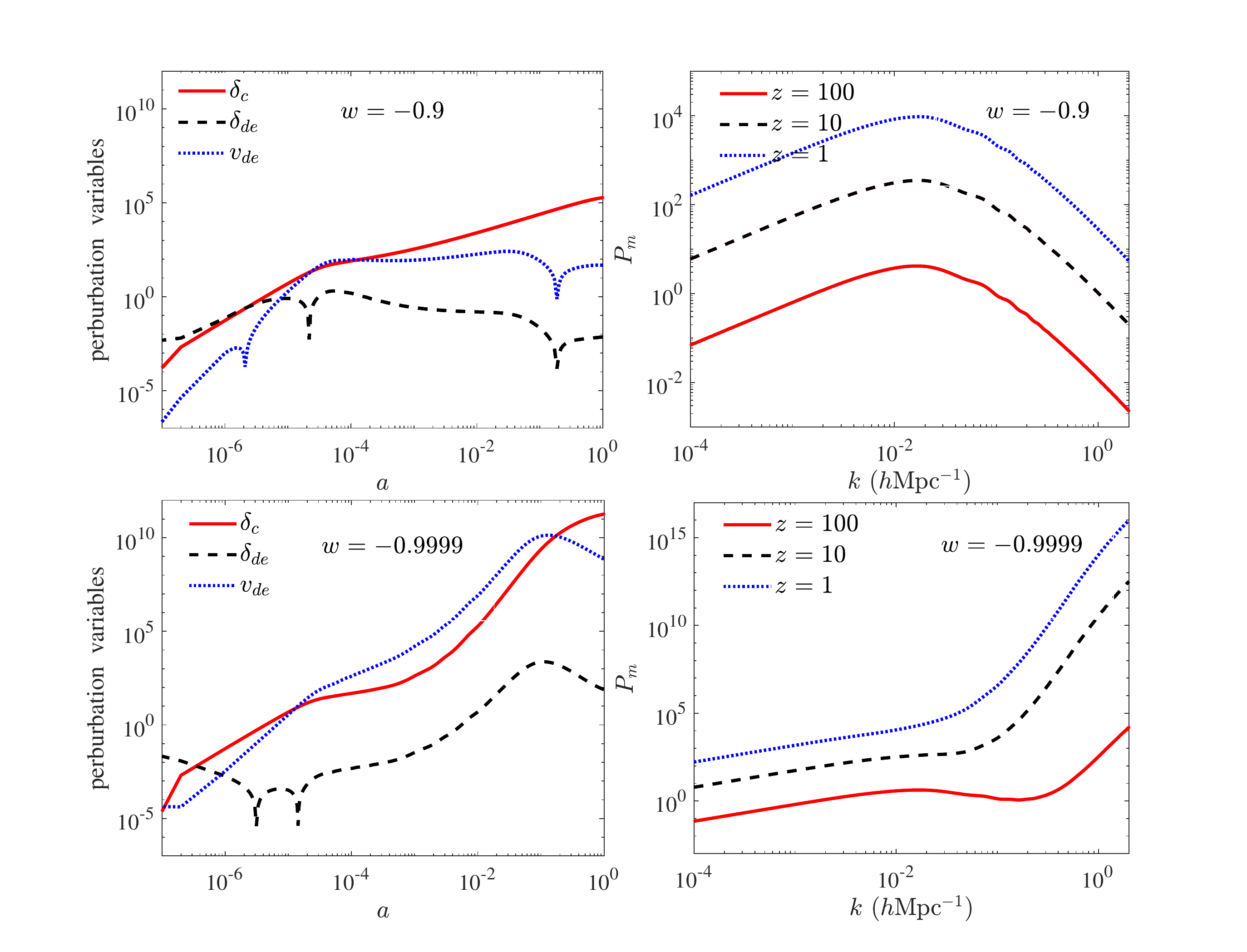}
\caption{Left panels: density evolutions of dark matter (red solid line) and dark energy (black dashed line), together with  velocity perturbations evolutions of dark energy (blue dotted line) at $k=1 \rm Mpc^{-1}$, $w=-0.9$ and  $w=-0.9999$. Right panels: matter power spectrum at different redshifts, $w=-0.9$ and  $w=-0.9999$, respectively. Here we use $Q_{de}^{\mu}=3\beta H\rho_c u_{de}^{\mu}$ with $\beta = -0.001$}
\label{fig2}
\end{figure*}

However, if the interacting dark energy model is expressed as $Q_{de}^{\mu}=3\beta H\rho_c u_{de}^{\mu}$, there is an explicit deviation of the dark matter velocity from that in the non-interacting case. The perturbed continuity and Euler equations Eq.\ref{DV} can be written as,
\begin{subequations}
\label{derho}
\begin{align}
&\delta_c' + kv_c + 3H_L' = -3aH\beta A, \\
&(v_c-B)'+\mathcal{H}(v_c-B) -kA = 3aH\beta (v_{c}-v_{de}) .
\end{align}
\end{subequations}
In this model, dark matter velocity $v_c$ is related to dark energy velocity $v_{de}$, which will affect the density perturbation of dark matter $\delta_c$. In the left side of Fig.\ref{fig2}, the evolutions of density and velocity perturbations are plotted at $k=1 \rm Mpc^{-1}$, $w=-0.9$ and $w=-0.9999$, respectively.

It can be seen from the figures that with $w$ approaching $-1$, the energy density and velocity perturbations would diverge. Predictably, this kind of interaction will affect the matter power spectrum. In the left side of Fig.\ref{fig2}, the matter power spectrum of different redshifts are shown when $w=-0.9$ and $w=-0.9999$.

\subsection{A visual attempt to  steer clear of the divergence}

Actually, if the interaction system does not contains $v_{de}$ directly, the divergence of $v_{de}$ would not cause physical problems. The previous transfer terms can be written as Eq.\ref{previous}, In that model, the energy transfer perturbation is proportional to $\rho_c$ and momentum transfer  perturbation is proportional to $\rho_c \times v_c$. However, this paper takes account of the stress-energy tensor $\delta T^\nu_{~\mu}$ instead of $\delta \rho$ and $v$. The energy transfer perturbation can be written as $Q_IA +\delta Q_I \propto \rho_J \delta_J+\rho_JA$, where $J$ is another fluid. In the mean time, the momentum transfer perturbation has the form $f_I + Q_I(v - B) \propto (\rho_J+p_J)(v_J-B)$. It can be figured out that $\rho_J \delta_J$ is $\delta T^0_{~0,J}$ component and $(\rho_J+p_J)v_J$ is $\delta T^i_{~0,J}$ component. So it is reasonable to set the interaction terms as,

\begin{subequations}
\label{rea}
\begin{align}
&Q_{de} = -Q_c = 3\beta H\rho_x\ ,\\
\begin{split}
&Q_{de}A + \delta Q_{de} =-Q_c A -\delta Q_c \\
&= 3\beta H (\delta \rho_x+ \rho_x A) \ ,\\
\end{split}
\\
\begin{split}
&Q_{de}(v-B)+f_{de}= -Q_{c}(v-B)-f_{c}\\
&=3\beta H(\rho_x+p_x)(v_x-B) \ ,
\end{split}
\end{align}
\end{subequations}
where the subscript $x$ refers to $c$ or $de$, which represents different kinds of interacting dark energy model. For simplicity, the energy transfer perturbation can be written as $\delta Q_{de} =-\delta Q_c = 3\beta H \delta \rho_x$.

Compared with previous models, the energy transfer rate in both background and perturbation are the same as the case above. And the momentum transfer perturbation makes correction of $(1+w)$. The interaction we propose modifies the total momentum transfer perturbation in a different way than done in the existing literature and implies that $f_c$ and $f_{de}$ are,
\begin{subequations}
\label{rea}
\begin{align}
&f_{de}= 3\beta H(\rho_x+p_x)(v_x-B) -Q_{de}(v-B),\\
&f_{c}=-3\beta H(\rho_x+p_x)(v_x-B)+Q_{c}(v-B) \ ,
\end{align}
\end{subequations}

\begin{figure}[tbp]
\includegraphics[scale=0.55]{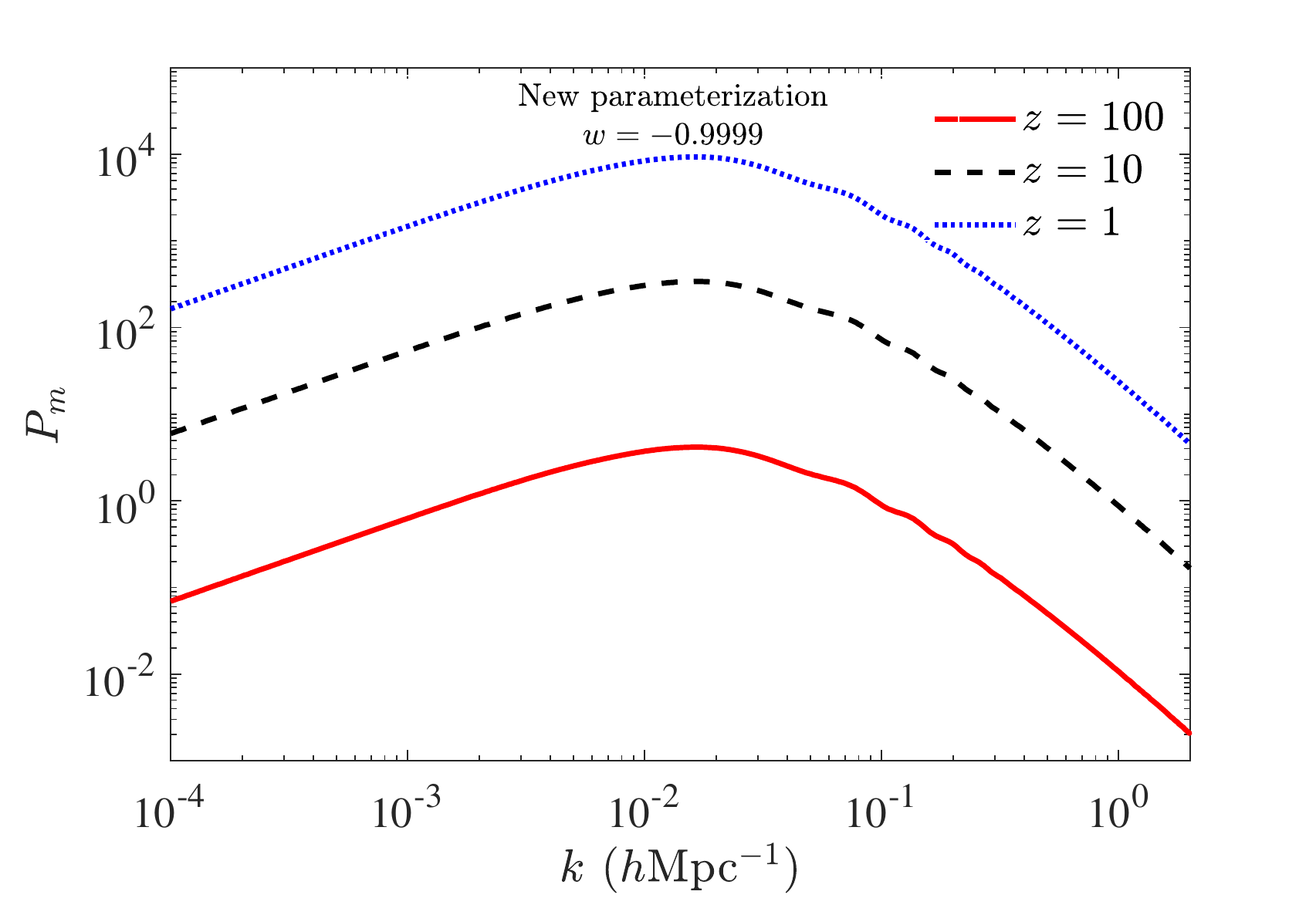}
\caption{Matter power spectrum using our new  parameterization, with $w=-0.9999$ and $\beta = -0.001$, the same conditions with lower right panel of Fig.\ref{fig2}.}
\label{fig9}
\end{figure}

When $x=c$, the perturbed continuity and Euler equations are just the same as $Q_{de}^{\mu}=3\beta H\rho_c u_{c}^{\mu}$, for $p_c = 0$. Thus the only imprint of the dark sector interaction on $\delta_c$ is via different background evolution of $\mathcal{H}$ and $\rho_c$.
As for $x =de$,  the perturbed continuity and Euler equations will be
\begin{subequations}
\label{derho}
\begin{align}
&\delta_c' + kv_c + 3H_L' = -\frac{3a\beta H \rho_{de} A}{\rho_c} , \\
\begin{split}
&(v_c-B)'+\mathcal{H}(v_c-B) -kA \\
&= \frac{3a\beta H \rho_{de}}{\rho_c}(v_c - B) - \frac{3(1+w)a\beta H \rho_{de}}{\rho_c}(v_{de}-B) .
\end{split}
\end{align}
\end{subequations}
Under this condition, when $v_{de}$ is needed in this system, there will be a correction of $(1+w)$, which can avoid the divergence efficaciously. Fig.\ref{fig9} shows the matter power spectrum when the parameterization is adopted with $w=-0.9999$. The result reveals it returns to the normal case.

Actually, we have no idea about the nature of dark matter and dark energy, much less the form of interaction between them. So there may be a number of different interacting models.  Based on the assumption in this paper, a more general parameterization can be put forward which would not diverge when $w$ is close to $-1$.
\begin{subequations}
\label{rea}
\begin{align}
&Q_{de} = -Q_c = C_1 \rho_c + C_2 \rho_{de} \ ,\\
&\delta Q_{de} = -\delta Q_c =  D_1\delta \rho_c+D_2\delta \rho_{de} \ ,\\
\begin{split}
&Q_{de}(v-B)+f_{de}= -Q_{c}(v-B)-f_{c}\\
&=E_1(\rho_c+p_c)(v_c-B)+ E_2(\rho_{de}+p_{de})(v_{de}-B)\ .
\end{split}
\end{align}
\end{subequations}

Set $C_1 = D_1 = E_1 = 3\beta H, C_2 = D_2 = E_2 = 0$, and the parameterization will be the same as $Q_{de}^{\mu}=3\beta H\rho_c u_{c}^{\mu}$. This parameterization will be effective to explore the interacting dark energy model in the PPF framework.

\section{Conclusion}
\label{sec5}
As the dominant components of the universe, dark energy may interact with cold dark matter in a direct, non-gravitational way. Since the interactions are very common in nature, the consideration of such interaction is quite reasonable. However, human beings have no idea about the specific formula of interaction. In order to testify the models with data, it is necessary to investigate the cosmological perturbations in the IDE scenario.

Dark energy is usually considered as non-adiabatic fluid. The perturbation equations are completed by defining the sound speed and adiabatic sound speed. Due to the incorrect treatment of the pressure perturbation of dark energy, the early time large-scale instability occurs in the IDE scenario. In order to solve this problem, the PPF framework was put forward to calculate the cosmological perturbations. Under this framework, there is no need to calculate the density and velocity perturbations, which can avoid the instability successfully. The fit results show that the full parameter space of this model can be explored.

Nevertheless, PPF framework cannot solve the problem completely in some specific models. For example, when $Q_{de}^{\mu}=3\beta H\rho_c u_{de}^{\mu}$, there is an explicit deviation of the dark matter velocity from that in the non-interacting case, which has an extra term proportion to $v_{de}$. With the PPF approach, $v_{de}$ diverges when $w$ is close to $-1$. So the perturbation of dark matter and the matter power spectrum will diverge.

In this paper, a visual attempt to steer clear of the divergence is put forward. Since the nature of interaction between dark matter and dark energy is unknown, the interaction terms can be written optionally as long as it is not ridiculous. So in this paper the density transfer and the momentum transfer are assumed to relate to stress-energy tensor $\delta T^0_{~0} = -\delta \rho $ and $\delta T^i_{~0} = -(\bar \rho+\bar p) v^i $, rather than $\delta \rho$ and $v_i$. Finally, the general parameterization is provided in which there is no divergence when $w$ is close to $-1$. The interaction between dark matter and dark energy might be further explored by using the general parameterization in the following works.

\section*{Acknowledgements}
J.-Q. Xia is supported by National Key R\&D Program of China No. 2017YFA0402600; the National Youth Thousand Talents Program and the National Science Foundation of China under grant No. 11422323, 11633001, and 11690023; and the Fundamental Research Funds for the Central Universities, grant No. 2017EYT01.

\end{document}